\documentclass[letterpaper,12pt,unknownkeysallowed, dvipsnames]{article} 
    \usepackage{algorithm}
    \usepackage[noend]{algpseudocode}
    \usepackage{booktabs}
    \usepackage{filemod}
	\usepackage[utf8]{inputenc}
	\usepackage[english]{babel}
	\usepackage{amsfonts}
	\usepackage{amssymb}
	\usepackage{enumitem}
	\usepackage{bbm}
	\usepackage{fullpage}
	\usepackage{amsmath}
	\usepackage{mathabx}
	\usepackage{mathtools}
    \usepackage{scalerel}
    \usepackage{xcolor}
    \usepackage{csquotes}
    \usepackage{array}
    \usepackage{tablefootnote}
    \usepackage{threeparttable}
    \newcolumntype{L}[1]{>{\raggedright\let\newline\\\arraybackslash\hspace{0pt}}m{#1}} 
    \usepackage{multirow}
    \mathtoolsset{showonlyrefs}

\usepackage[round,longnamesfirst]{natbib}
\renewcommand{\appendix}
{\parindent 0cm\parskip 
5pt\setcounter{equation}{0}
\setcounter{section}{0}
\renewcommand{\thesection}{A.\arabic{section}}
\renewcommand{\theequation}{A.\arabic{equation}}
\setcounter{lemma}{0}\renewcommand{\thelemma}{A.\arabic{lemma}}
\setcounter{theorem}{0}\renewcommand{\thetheorem}{A.\arabic{theorem}}
}

\makeatletter
\def\blfootnote{\xdef\@thefnmark{}\@footnotetext}
\makeatother

\makeatletter
\def\@sect#1#2#3#4#5#6[#7]#8{\ifnum #2>\c@secnumdepth
     \let\@svsec\@empty\else
     \refstepcounter{#1}\edef\@svsec{\csname the#1\endcsname. \hskip 0.4em}\fi
     \@tempskipa #5\relax
      \ifdim \@tempskipa>\z@
        \begingroup #6\relax
          \@hangfrom{\hskip #3\relax\@svsec}{\interlinepenalty \@M #8\par}
        \endgroup
       \csname #1mark\endcsname{#7}\addcontentsline
         {toc}{#1}{\ifnum #2>\c@secnumdepth \else
                      \protect\numberline{\csname the#1\endcsname}\fi
                    #7}\else
        \def\@svsechd{#6\hskip #3\relax  
                   \@svsec #8\csname #1mark\endcsname
                      {#7}\addcontentsline
                           {toc}{#1}{\ifnum #2>\c@secnumdepth \else
                             \protect\numberline{\csname the#1\endcsname}\fi
                       #7}}\fi
     \@xsect{#5}}
\makeatother

\makeatletter
\renewcommand{\section}{\@startsection{section}{1}{0mm}{-\baselineskip}{0.25\baselineskip}{\centering\normalfont\normalsize\bf}}
\renewcommand{\subsection}{\@startsection{subsection}{2}{0mm}{-\baselineskip}{0.25\baselineskip}{\raggedright\normalfont\normalsize\bf}}
\renewcommand{\subsubsection}{\@startsection{subsubsection}{3}{0mm}{-\baselineskip}{0.25\baselineskip}{\raggedright\itshape\small}}
\def\@begintheorem#1#2{\trivlist \item[\hskip \labelsep{\bf #1\ #2}]\it}
\makeatother



\renewcommand{\thesection}{\arabic{section}}

 \newcommand*\mwidebar[1]{%
   \hbox{%
     \vbox{%
       \hrule height 0.5pt %
       \kern0.30ex%         % 
       \hbox{%
         \kern-0.02em%      % 
         \ensuremath{#1}%
         \kern-0.02em%      % 
       }%
     }%
   }%
} 

\newcommand*\lowwidebar[1]{%
   \hbox{%
     \vbox{%
       \hrule height 0.5pt % 
       \kern0.20ex%         % 
       \hbox{%
         \kern-0.02em%      % 
         \ensuremath{#1}%
         \kern-0.02em%      % 
       }%
     }%
   }%
}

\usepackage{tikz,amsmath,environ}
\usetikzlibrary{decorations.pathreplacing,calc}
\tikzset{
    ncbar angle/.initial=90,
    ncbar/.style={
        to path=(\tikztostart)
        -- ($(\tikztostart)!#1!\pgfkeysvalueof{/tikz/ncbar angle}:(\tikztotarget)$)
        -- ($(\tikztotarget)!($(\tikztostart)!#1!\pgfkeysvalueof{/tikz/ncbar angle}:(\tikztotarget)$)!\pgfkeysvalueof{/tikz/ncbar angle}:(\tikztostart)$)
        -- (\tikztotarget)
    },
    ncbar/.default=0.5cm,
    freaky dim/.default=4pt,
    freaky/.style={
        to path={let \p1=(\tikztostart),\p2=(\tikztotarget) in (\tikztostart)
        -- ($(\tikztostart)!#1!\pgfkeysvalueof{/tikz/ncbar angle}:(\tikztotarget)$)
        -- ({\x1-#1},{(\y1+\y2)/2-abs(#1)})
        -- ({\x1-2*#1},{(\y1+\y2)/2})
        -- ({\x1-#1},{(\y1+\y2)/2+abs(#1)})
        -- ($(\tikztotarget)!($(\tikztostart)!#1!\pgfkeysvalueof{/tikz/ncbar angle}:(\tikztotarget)$)!\pgfkeysvalueof{/tikz/ncbar angle}:(\tikztostart)$)
        -- (\tikztotarget)}
    },
    freaky/.default=0.5cm,
}

\tikzset{freaky left brace/.style={freaky=0.5ex}}
\tikzset{freaky right brace/.style={freaky=-0.5ex}}

\tikzset{square left brace/.style={ncbar=0.5ex}}
\tikzset{square right brace/.style={ncbar=-0.5ex}}

\tikzset{round left paren/.style={ncbar=0.3cm,out=115,in=-115}}
\tikzset{round right paren/.style={ncbar=0.3cm,out=65,in=-65}}

\NewEnviron{flcases}{\setbox0=\hbox{$\,\begin{matrix}\BODY\end{matrix}\,$}%
  \setbox2=\hbox{\begin{tikzpicture}
    \draw [thick] (0,\botdim) to [freaky left brace] (0,\topdim);
    \copy0
  \end{tikzpicture}}
  \vcenter{\hbox{\copy2}}%
}

\def\topdim{\the\dimexpr+\ht0+.5\ht\strutbox-.5\dp\strutbox-3pt\relax}
\def\botdim{\the\dimexpr-\ht0+.5\ht\strutbox-.5\dp\strutbox+3pt\relax}

\allowdisplaybreaks

\usepackage{fancyheadings}
\begin{document}
\vspace*{0.2cm}

\blfootnote{\hspace*{-0.25in}Alberto Abadie, Department of Economics, MIT, abadie@mit.edu.~Anish Agarwal, Department
of Industrial Engineering and Operations Research, Columbia University, aa5194@columbia.edu. Guido Imbens, Graduate School of Business, Stanford, imbens@stanford.edu. Siwei Jia, Amazon.com, siweijia@amazon.com. James McQueen, Amazon.com, jmcq@amazon.com. Serguei Stepaniants, Amazon.com,  sergueis@amazon.com. Santiago Torres, MIT, storresp@mit.edu. We are grateful to Haruki Kono, Soonwoo Kwon, Vira Semenova and seminar participants at Amazon.com, Brandeis, Princeton, and Toronto for helpful comments. %and to Brad Efron for making a subset of the Cochrane data available to us. 
The views expressed in this article are solely those of the authors and do not necessarily reflect those of Amazon.com. This research is partially supported by ONR grants N00014-24-1-2687 (Abadie) and N00014-19-1-2468 (Imbens).}
\vskip 20pt
\centerline{\large\bf Estimating the Value of Evidence-Based Decision Making}
  \begin{center}
  
    \vskip 10pt
    
    {\large
     \lineskip .5em
   \begin{tabular}[t]{ccccccc}
       Alberto Abadie&&Anish Agarwal&&Guido Imbens&&Siwei Jia\\[.2ex]
       MIT&&Columbia&&Stanford&&Amazon.com\\[1ex]
    \end{tabular} 
    \begin{tabular}[t]{ccccc}
       James McQueen&&Serguei Stepaniants&& Santiago Torres\\[.2ex]
       Amazon.com&&Amazon.com&&MIT\\[1ex]
    \end{tabular} 
      \par}
      \vskip 1em
      {\large \today} \par
       \vskip 1em
  \end{center}\par

\bigskip
\begin{center}\normalsize\bf\text{Abstract}
 \end{center}\begin{quote}\normalsize
\noindent{In an era of data abundance, statistical evidence is increasingly critical for business and policy decisions. Yet, organizations lack empirical tools to assess the value of evidence-based decision making (EBDM), optimize statistical precision, and balance the costs of evidence-gathering strategies against their benefits. To tackle these challenges, this article introduces an empirical framework to estimate the value of EBDM and evaluate the return on investment in statistical precision and project ideation. The framework leverages parametric and nonparametric empirical Bayes methods to account for parameter heterogeneity and measure how statistical precision changes the value of evidence. The value extracted from statistical evidence depends critically on how organizations translate evidence into policy decisions. Commonly used decision rules based on statistical significance can leave substantial value unrealized and, in some cases, generate negative expected value.}
\end{quote}

\addtolength{\baselineskip}{0.5\baselineskip}

\newpage
 
\section{Introduction} \label{sec:introduction}

Many organizations use randomized experiments and observational studies to improve their decision making. For example, \citet{gupta2019top} write ``Together [Airbnb, Amazon, Booking.com, Facebook, Google, LinkedIn, Lyft, Microsoft, Netflix, Twitter, Uber, Yandex, and Stanford University] have tested more than one hundred thousand experiment treatments last year.'' The fact that so many organizations conduct such a large number of experiments suggests that these organizations believe that data evidence provides significant value in guiding business and policy decisions. However, we are unaware of empirical tools that organizations can use to assess the actual value of their EBDM practices. In the absence of such tools, it is difficult to determine whether too much experimentation is being conducted or too little, whether experiments are too large or too small, and whether the right experiments are being undertaken. Part of the challenge in evaluating the value of EBDM lies in the need to describe the role of evidence in the business and policy decision-making process. In other words, estimating the value of EBDM requires assumptions about what organizations will do with and without various amounts of evidence, which they can choose to generate at some cost.

In this article, we propose an empirical Bayes estimator for the value of EBDM. We study the problem of a decision maker choosing whether to adopt a particular policy intervention. We use the term ``agent'' to refer to the decision maker, and ``policy,'' ``intervention,'' and ``treatment'' interchangeably to refer to the policy intervention under scrutiny. The agent can implement the intervention based on prior information or gather additional information at some cost—for example, by running an experimental or observational evaluation of the intervention's effect. At the stage where the agent decides whether to implement the intervention, they aim to maximize utility based on the available information. We derive expressions for the value of additional information and demonstrate how to estimate this value using metadata on estimates of the effects of business and policy interventions, along with their standard errors.

Our framework allows decision makers to assess in a principled way the value of experimental and non-experimental studies, and how design choices affect that value. Currently, many organizations decide on the precision of their studies based on power calculations. These  do not take into account the costs and benefits of  EBDM and instead rely on statistical conventions (for example, requiring 80\% power for tests at the 5\% level). Additionally, our framework enables decision makers to {\it ex ante} assess whether an experiment is worth conducting based on its cost and expected benefits (with benefits increasing with the decision maker's initial uncertainty about the intervention’s effect). 
\medskip

\noindent{\em Related literature.}---This article builds on foundational work by \cite{blackwell1951experiments} and \cite{howard1966voi} on the value of information. It extends that framework to the applied domain of EBDM, combining costs, precision, and empirical Bayes estimation into a flexible tool for real-world decisions and counterfactual analysis. The resulting methods are particularly suited to data-rich environments, where organizations must balance the benefits of additional information against the costs of generating it.

The empirical setting is also closely related to that of meta-analytic studies \citep{hedges1985statistical, higgins2019cochrane} in that it leverages information from many individual studies. However, unlike meta-analysis, the goal of this article is not to assess the effectiveness of a set of policies but to quantify the value brought by empirical evidence in guiding better policy decisions.

A key component of the EBDM estimand is the expected value of the positive part of the predicted policy payoff, that is, the expectation of the maximum of the predicted policy payoff and zero. A related but distinct object is considered in \cite{semenova2023generalized} in the context of estimating the size of a latent population whose outcomes are observed regardless of treatment exposure. The estimand in \cite{semenova2023generalized} targets the distribution of an expectation given observed covariates, and does so in a single study setting. In contrast, our estimand pertains to the distribution of predicted policy payoffs across many studies. 

This article also contributes to the growing literature on empirical Bayes methods \citep[see][for foundational work on empirical Bayes]{morris1983parametric,Efron_2010}. Empirical Bayes and related shrinkage techniques play a central role in applied economics, informing research on teacher and school value-added \citep{chetty2014measuring, angrist2017leveraging}, neighborhood effects \citep{Chetty2018impact}, income dynamics \citep{gu2017unobserved}, and racial discrimination \citep{kline2024discrimination}, among other topics. As empirical Bayes methods gain traction in applied economics, a parallel methodological literature has emerged in econometrics \citep{koenker2014convex, abadie2019choosing, fessler2019how, armstrong2022robust, kwon2023optimal, koenker2024empirical, chen2024empirical}. \cite{WALTERS2024183} provides a comprehensive account of empirical Bayes methods and their applications in economics. This article applies both parametric and nonparametric empirical Bayes techniques to estimate the distribution of policy payoffs in settings where the data contain information about the effects of many policies. 
 
\section{The value of EBDM}
The notation $X \sim (\theta, \sigma^2)$ indicates that the random variable $X$ has mean $\theta$ and variance $\sigma^2$. When $X$ follows a Gaussian distribution with mean $\theta$ and variance $\sigma^2$, it is denoted as $X \sim N(\theta, \sigma^2)$. $f_X(\cdot)$ represents a probability density function of the random variable $X$, while $f_{X|W}(\cdot|w)$ denotes a conditional probability density function of $X$ given $W = w$. $F_X(\cdot)$ denotes the cumulative distribution function of $X$, and $F_{X\mid W}(\cdot \mid w)$ is the cumulative distribution function of  $X$ given $W = w$. The functions $\phi(\cdot)$ and $\Phi(\cdot)$ refer to the probability density function and cumulative distribution function of the standard Gaussian distribution, respectively.

\subsection{Setup}
Consider the problem of a risk-neutral decision maker tasked with choosing whether to adopt a particular policy for a population of units. 
%If the decision maker is not risk neutral, the analysis still applies with payoffs expressed in terms of utility.
The \emph{ex ante} unknown per-unit payoff of the policy, $\tau$, follows a distribution with known probability density function $f_\tau(\cdot)$ and mean $\mu=E[\tau]$. In an organization where teams generate new ideas for policies or interventions, $f_\tau(\cdot)$ can be thought of as the distribution of the quality of those ideas. For retrospective estimation tasks, we take this distribution as fixed. However, organizations can shift it, for example, by prioritizing high-risk projects with substantial upside, protecting agents from failure, or allocating resources to exploratory projects.

In the absence of additional  information, the agent launches the policy if the expected payoff from launching is positive,
\[
\mu-c_L> 0,
\]
where $c_L$ is the cost of launching per-unit. The expected value of this decision is
$\max\{\mu-c_L,0\}$.

Now, suppose the agent has the option to obtain additional information about the policy payoff at some cost. Specifically, the agent can acquire a signal, $\widehat{\tau}$, distributed as
\begin{equation}
\widehat\tau\,|\,\tau, \sigma^2 \sim N(\tau,\sigma^2),
\label{equation:GaussAprox}
\end{equation}
at a cost of $c_F + c(\sigma^2)$, where $c_F \geq 0$, $c(\cdot) \geq 0$, and $c'(\cdot) \leq 0$. This setup captures the information obtained from studies estimating policy effects using experimental or observational data. The constant $c_F$ reflects the fixed cost of conducting a data-driven policy evaluation, while the function $c(\cdot)$ captures the cost of precision, which partly depends on the study's sample size. The restriction on the derivative $c'(\cdot)$ indicates that obtaining more precise information is weakly more expensive. The assumption of Gaussianity for the distribution of $\widehat\tau\,|\,\tau, \sigma^2$ is motivated by the approximate Gaussian nature of the large-sample distributions of many commonly used estimators of treatment effects.

After observing the signal $\widehat\tau$, the expected payoff of the policy is 
\begin{align*}
    E[\tau|\widehat\tau=t] &= \int u f_{\tau|\widehat\tau}(u|t) du\\
                         &= \frac{\displaystyle\int \frac{u}{\sigma}\phi((t-u)/\sigma)f_{\tau}(u) du}{\displaystyle\int \frac{1}{\sigma}\phi((t-u)/\sigma)f_\tau(u) du}.
\end{align*}
If a signal is observed, the agent launches the policy if
\[
E[\tau|\widehat\tau] - c_L> 0.
\]

For any set $\mathcal A$, let $I_{\mathcal A}(x)$ be the function that takes value one if $x\in \mathcal A$, and value zero otherwise. The expected payoff with EBDM for a fixed value of the variance of the signal $\sigma^2$ is 
\begin{align}
V(\sigma^2)
&=E\left[I_{(0,\infty)}(E[\tau|\widehat\tau] - c_L)(\tau - c_L) \right]\nonumber\\
&=E\left[I_{(0,\infty)}(E[\tau|\widehat\tau] - c_L)(E[\tau|\widehat\tau]-c_L)\right]\nonumber\\
&=E\left[\max\left\{E[\tau|\widehat\tau]-c_L,0\right\}\right].
\end{align}

Define $V(\infty)$ as the expected payoff with no information beyond the distribution of $\tau$, that is,
\[
V(\infty) = \max\{\mu-c_L,0\}.
\]
Because $\max\{x,0\}$ is a convex function of $x$, Jensen's inequality implies,
\[
\text{V}(\sigma^2)\geq \max\{\mu-c_L,0\}=V(\infty).
\]
The value of evidence (\text{VoE}) is the difference in expected payoffs $V(\sigma^2)$ and $V(\infty)$, which is nonnegative, minus the cost of acquiring the information, which is generally positive:
\begin{align}\label{eq:VoE}
\text{VoE}(\sigma^2) = V(\sigma^2) - V(\infty) - (c_F + c(\sigma^2)).
\end{align}
A second version of $\text{VoE}$, which we term $\text{VoID}$ (for Value of Information under Default adoption) is obtained when, in the absence of additional information about the effect of the intervention, the intervention is always deployed and so $V(\infty)=\mu -c_L$: 
\begin{align}\label{eq:VoEL}
\text{VoID}(\sigma^2) = V(\sigma^2) - (\mu -c_L) - (c_F + c(\sigma^2)).
\end{align}
$\text{VoID}(\sigma^2)$ is motivated by settings with ex-ante (pre-evaluation) ambiguity on the distribution of $\tau$, and agents who have a bias for action in the presence of such ambiguity.

\subsection{A motivating example}

A common instance of the setting described above is one where the decision maker obtains experimental evidence on the effect of the policy. Consider an experiment with $N$ units: $i=1, \ldots, N$. The experimenter assigns $N_1$ units at random to treatment and the remaining $N_0=N-N_1$ to control. If unit $i$ is treated, an outcome is drawn
\begin{align*}
Y_i(1) \sim (\theta_1, \sigma_1^2).
\end{align*}
If unit $i$ is untreated, the outcome is drawn 
\begin{align*}
Y_i(0) \sim (\theta_0, \sigma_0^2).
\end{align*}
Let $W_i$ be an indicator of treatment for unit $i$. We observe $Y_i=Y_i(1)W_i+Y_i(0)(1-W_i)$. 
The average effect of the treatment is
\begin{align*}
    \tau=\theta_1-\theta_0.
\end{align*} 
A simple estimator of $\tau$ is the difference in mean outcomes between treated and nontreated, 
\begin{align*}
\widehat{\tau} = \frac{1}{N_1}\sum^{N}_{i = 1} W_i Y_i - \frac{1}{N_0}\sum^{N}_{i = 1} (1-W_i)Y_i. 
\end{align*}
Then, for large $N_0$ and $N_1$, equation \eqref{equation:GaussAprox} holds approximately, with
\begin{align*}
\sigma^2 = \frac{\sigma_1^2}{N_1} + \frac{\sigma_0^2}{N_0}. \label{eq:variance_defn}
\end{align*}
When a fraction $p=N_1/N$ of units are assigned to treatment, and assuming that the only variable cost of the experiment comes from recruiting subjects at a cost $\kappa$ per subject, the total cost of the experiment 
is 
\[
c_F + \frac{\kappa}{\sigma^2} \left(\frac{\sigma_1^2}{p} + \frac{\sigma_0^2}{1-p}\right),
\]
where $c_F$ is the fixed cost of the experiment.

Neyman's allocation rule, $p/(1-p) = \sigma_1/\sigma_0$, 
minimizes the variance $\sigma^2$ for a fixed total number of experimental subjects, $N$. When this rule is applied to allocate subjects between a treatment and a control group, the cost of the experiment becomes 
\[
c_F + \kappa \frac{(\sigma_1 + \sigma_0)^2}{\sigma^2}.
\]

%In this example, as in the rest of the article, $\tau$ is the unit level policy effect, and $c_L$ and $c(\sigma^2)$ have a per-unit interpretation.  

In some settings, researchers favor treatment effect parameters free of units of measurement, such as lift $\tau = (\theta_1-\theta_0)/\theta_0$. Let
\[
\widehat\tau = \frac{\displaystyle\frac{1}{N_1}\displaystyle\sum^{N}_{i = 1} W_i Y_i - \displaystyle\frac{1}{N_0}\displaystyle\sum^{N}_{i = 1} (1-W_i)Y_i}{\displaystyle\frac{1}{N_0}\displaystyle\sum^{N}_{i = 1} (1-W_i)Y_i}. 
\]
Then, for large $N_0$ and $N_1$, equation \eqref{equation:GaussAprox} holds with
\[
\sigma^2 = \frac{1}{\theta_0^2}\left(\frac{\sigma_1^2}{N_1}+(1+\tau)^2\frac{\sigma_0^2}{N_0}\right).
\]
For values of the lift parameter close to zero, as is common in many online experimentation settings, we can approximate
\[
\sigma^2 \approx \frac{1}{\theta_0^2}\left(\frac{\sigma_1^2}{N_1}+\frac{\sigma_0^2}{N_0}\right).
\]
In this case, Neyman allocation yields, 
\[
c_F + \kappa\frac{(\sigma_1 + \sigma_0)^2}{\theta_0^2\sigma^2},
\]

\subsection{A Gaussian distribution for $\tau$}
\label{section:gaussian_prior}

This section derives a simple closed-form expression for $V(\sigma^2)$ under the assumption that the distribution of $\tau$ is Gaussian,
\begin{align}
\tau\sim N(\mu, \gamma^2).
\label{equation:tau_sigma}
\end{align}
While equation \eqref{equation:GaussAprox} is supported by the Central Limit Theorem in studies with large samples, equation \eqref{equation:tau_sigma} imposes two important restrictions. First, $\tau$ is a Gaussian random variable. Second, implicit in the notation is the assumption that the distribution of $\tau$ is independent of $\sigma^2$. The first is a strong parametric restriction. 
The Gaussian approximation for $\tau$ could be valid in some settings but questionable in others. The second restriction could be violated, for example, if researchers adapt the power of individual studies to take into account prior information about the effect on the treatment. We dispose of these two restrictions later in the article. We adopt them in this section, however, to obtain closed-form formulas for the value of EBDM in a simple setting.

If equations \eqref{equation:GaussAprox} and \eqref{equation:tau_sigma} hold, the marginal distribution of $\widehat{\tau}$ is $\widehat\tau\sim N(\mu, \gamma^2 + \sigma^2)$.
The posterior for $\tau$ is given by
\begin{align*}
\tau \,|\, \widehat{\tau} \sim N\left(
\frac{\mu / \gamma^2 + \widehat{\tau} / \sigma^2}{1 / \gamma^2 + 1 /  \sigma^2}, 
\frac{1}{1 / \gamma^2 + 1 / \sigma^2}
\right).
\end{align*}
The expected payoff with EBDM is 
\begin{align*}
V(\sigma^2) 
&=E\left[\max\left\{\frac{\mu / \gamma^2 + \widehat\tau / \sigma^2}{1 / \gamma^2 + 1 /  \sigma^2}-c_L,0\right\}\right].
\end{align*}
Let 
\[
Z= \frac{\mu / \gamma^2 + \widehat\tau / \sigma^2}{1 / \gamma^2 + 1 /  \sigma^2}-c_L.
\]
Recall that the marginal distribution of $\widehat\tau$ is Gaussian with mean $\mu$ and variance $\gamma^2+\sigma^2$. As a result,
\begin{equation}
Z\sim N \left(\mu-c_L,\frac{\gamma^4}{\gamma^2+\sigma^2}\right).
\label{equation:distribution_Z}
\end{equation}
Now, $V(\sigma^2)$ is the the first moment of the Gaussian distribution in \eqref{equation:distribution_Z} censored from below at zero, 
\begin{align}
V(\sigma^2) 
&=(\mu - c_L) \Phi\left(\frac{\mu - c_L}{\gamma^2/\sqrt{\gamma^2+\sigma^2}} \right) + \frac{\gamma^2}{\sqrt{\gamma^2+\sigma^2}}\phi\left(\frac{\mu - c_L}{\gamma^2/\sqrt{\gamma^2+\sigma^2}}\right).
\label{equation:Vsigma}
\end{align}
The derivatives of $V(\sigma^2)$ with respect to $\sigma^2$, $\gamma^2$, and $\mu$ are
\begin{align}\frac{\partial V(\sigma^2)}{\partial \sigma^2} 
&= - \frac{\gamma^2}{2(\gamma^2+\sigma^2)^{3/2}}\phi\left(\frac{\mu-c_L}{\gamma^2/\sqrt{\gamma^2+\sigma^2}}\right)\leq 0,\\
\frac{\partial V(\sigma^2)}{\partial \gamma^2} &= \frac{\gamma^2+2\sigma^2}{2(\gamma^2+\sigma^2)^{3/2}}\phi\left(\frac{\mu-c_L}{\gamma^2/\sqrt{\gamma^2+\sigma^2}}\right)\geq 0,
\shortintertext{and}
\frac{\partial V(\sigma^2)}{\partial \mu} &= \Phi\left(\frac{\mu-c_L}{\gamma^2/\sqrt{\gamma^2+\sigma^2}}\right)\geq 0.
\end{align}
Higher precision of the signal $\widehat\tau|\tau$ increases the expected payoff from EBDM. In addition, the value of experimentation increases when an organization increases the variance of the distribution of true effects---that is, the variance of idea quality. The derivative of $V$ with respect to $\mu$ implies
\begin{align}
\frac{\partial\text{VoE}}{\partial \mu} &= \Phi\left(\frac{\mu-c_L}{\gamma^2/\sqrt{\gamma^2+\sigma^2}}\right) - I_{(0,\infty)}(\mu-c_L),
\shortintertext{and}
\frac{\partial\text{VoID}}{\partial \mu} &= \Phi\left(\frac{\mu-c_L}{\gamma^2/\sqrt{\gamma^2+\sigma^2}}\right) - 1.
\end{align}
$\text{VoE}$ peaks at $\mu = c_L$ and decreases monotonically with $|\mu - c_L|$. When $|\mu - c_L|$ is large, a simple rule that selects policies based solely on the sign of $\mu - c_L$ gets most decisions right, leaving little room for additional evidence to add value. $\text{VoID}$, the value of EBDM when policies are adopted by default in the absence of additional information, decreases monotonically with $\mu$. $\text{VoID}$ is particularly large when the distribution of $\tau$ is concentrated on negative values, as additional information winnows out many ineffective or counterproductive policies.

Notice that
\begin{equation*}
\lim_{\sigma^2\rightarrow \infty} V(\sigma^2) = \max\{\mu-c_L,0\}=V(\infty)
\end{equation*}
and 
\[
\lim_{\sigma^2\rightarrow 0} V(\sigma^2) =
(\mu -c_L)\Phi\left(\frac{\mu-c_L}{\gamma}\right)+\gamma\phi\left(\frac{\mu-c_L}{\gamma}\right).\vspace*{.2cm}
\]
As $\sigma^2\rightarrow\infty$, we lose any additional information about the value of $\tau$ beyond its distribution, and $V(\sigma^2)$ converges to $V(\infty)$. As $\sigma^2\rightarrow 0$, the information gathering process reveals the value of $\tau$. In this case, $V(\sigma^2)$ converges to $E[\max\{\tau - c_L, 0\}]$, the mean of the distribution of $\tau - c_L$ censored at zero.

So far, we have treated $\sigma^2$ as a constant. We now allow $\sigma^2$ to have a  non-degenerate distribution, independent of $\tau$.
In this case, the average payoff of EBDM is
\begin{align}
 V&=E\Big[\max\{E\big[\tau|\widehat\tau,\sigma\big]-c_L,0\}\Big]\nonumber\\&=E\Big[E\big[\max\{E[\tau|\widehat\tau,\sigma]-c_L,0\}|\sigma\big]\Big]\nonumber\\
    &=E\left[(\mu - c_L) \Phi\left(\frac{\mu - c_L}{\gamma^2/\sqrt{\gamma^2+\sigma^2}} \right) + \frac{\gamma^2}{\sqrt{\gamma^2+\sigma^2}}\phi\left(\frac{\mu - c_L}{\gamma^2/\sqrt{\gamma^2+\sigma^2}}\right)\right],
    \label{equation:V}
\end{align}
with the expectation taken over the distribution of $\sigma^2$.

\subsection{A Gaussian mixture distribution for $\tau$}
\label{section:mixture_prior}

In Section \ref{section:Simulations} we use a Gaussian mixture distribution to evaluate the effects of misspecification of the distribution of $\tau$ on EBDM value estimates. Suppose $\tau$ follows a mixture of $k$ Gaussian distributions with parameters $(\mu_1,\gamma^2_1), \ldots, (\mu_k,\gamma^2_k)$, and mixture probabilities $p_1, \ldots, p_k$. Let $\widehat\tau=\tau + \varepsilon$, where $\varepsilon$ is independent Gaussian noise with variance $\sigma^2$. Conditional on $\tau\sim N(\mu_j,\gamma^2_j)$, we have
\[
E[\tau|\widehat\tau, \tau\sim N(\mu_j,\gamma^2_j)] = \frac{\mu_j / \gamma_j^2 + \widehat{\tau} / \sigma^2}{1 / \gamma_j^2 + 1 /  \sigma^2}.
\]
As a result,
\begin{align}
E[\tau|\widehat\tau] &= \sum_{j=1}^k E[\tau|\widehat\tau, \tau\sim N(\mu_j,\gamma^2_j)]
\Pr(\tau\sim N(\mu_j,\gamma^2_j)|\widehat\tau)\\
&=\frac{\displaystyle\sum_{j=1}^k\left(\displaystyle\frac{\mu_j / \gamma^2_j + \widehat{\tau} / \sigma^2}{1 / \gamma^2_j + 1 /  \sigma^2}\right)\frac{1}{\sqrt{\gamma_j^2+\sigma^2}}\phi\left(\displaystyle\frac{\widehat\tau-\mu_j}{\sqrt{\gamma^2_j+\sigma^2}}\right)p_j}{\displaystyle\sum_{j=1}^k\frac{1}{\sqrt{\gamma_j^2+\sigma^2}}\phi\left(\displaystyle\frac{\widehat\tau-\mu_j}{\sqrt{\gamma^2_j+\sigma^2}}\right)p_j}.
\label{equation:mixture_posterior_mean}
\end{align}
The simulations in Section \ref{section:Simulations} use the Gaussian mixture model to capture deviations from normality in the distribution of $\tau$.

\section{Empirical Bayes estimation}
\label{section:empirical_Bayes}
In this section, we analyze a setting with $n$ realizations from the distribution of $(\tau,\sigma,\widehat\tau,\widehat\sigma)$, where only the estimates $\widehat\tau$ and $\widehat\sigma$ are observed. 
We use the observations on  $\widehat\tau$ and $\widehat\sigma$  to estimate $\mu$ and $\gamma^2$ using an Empirical Bayes strategy. 
$(\widehat\tau,\widehat\sigma)$ represent point estimates and their corresponding standard errors for a set of policy evaluations in the dataset. We examine both the homoskedastic case, where $\sigma^2$ is constant, and the heteroskedastic case, where $\mbox{var}(\sigma^2)>0$. Throughout our analysis, we approximate the per-unit launch cost as $c_L\approx 0$. Alternatively, we can interpret $\tau$ as representing the net benefits of the treatment after accounting for the launch cost.

We employ a database of thousands of online experiments run by Upworthy to illustrate the applicability of our methods. Upworthy is a U.S. online news and media publisher that built a large following in the 2010s by pairing positive, uplifting stories with optimized headline-and-image packages designed to drive clicks and shares. Upworthy pioneered large-scale A/B testing of these packages, routinely randomizing visitors across alternative headlines and images for the same article preview and using click-through performance to guide editorial and distribution choices.

The Upworthy Research Archive records thousands of randomized experiments run by Upworthy between January 24, 2013 and April 30, 2015. We restrict our analysis to the Upworthy Exploratory Dataset, which contains outcomes for $4{,}873$ online experiments. Each experiment compared alternative headline-and-image packages for the same article preview.  We analyze outcomes for the first two packages deployed in each experiment, labeling the first as the control arm and the second as the treatment arm. 
For each package, the archive reports impressions and clicks. We remove from the sample all experiments with fewer than 100 impressions in one of the experimental arms, which yields a sample of $n=4{,}857$ online experiments. We then compute each experimental arm’s click rate per thousand impressions and define $\widehat{\tau}$ as the treatment–control difference in those rates. We also compute the standard error of $\widehat{\tau}$.

\begin{figure}[!tp]
    \centering
    \caption{Distribution of estimated treatment effects in the Upworthy data}
    \includegraphics[width=0.9\linewidth]{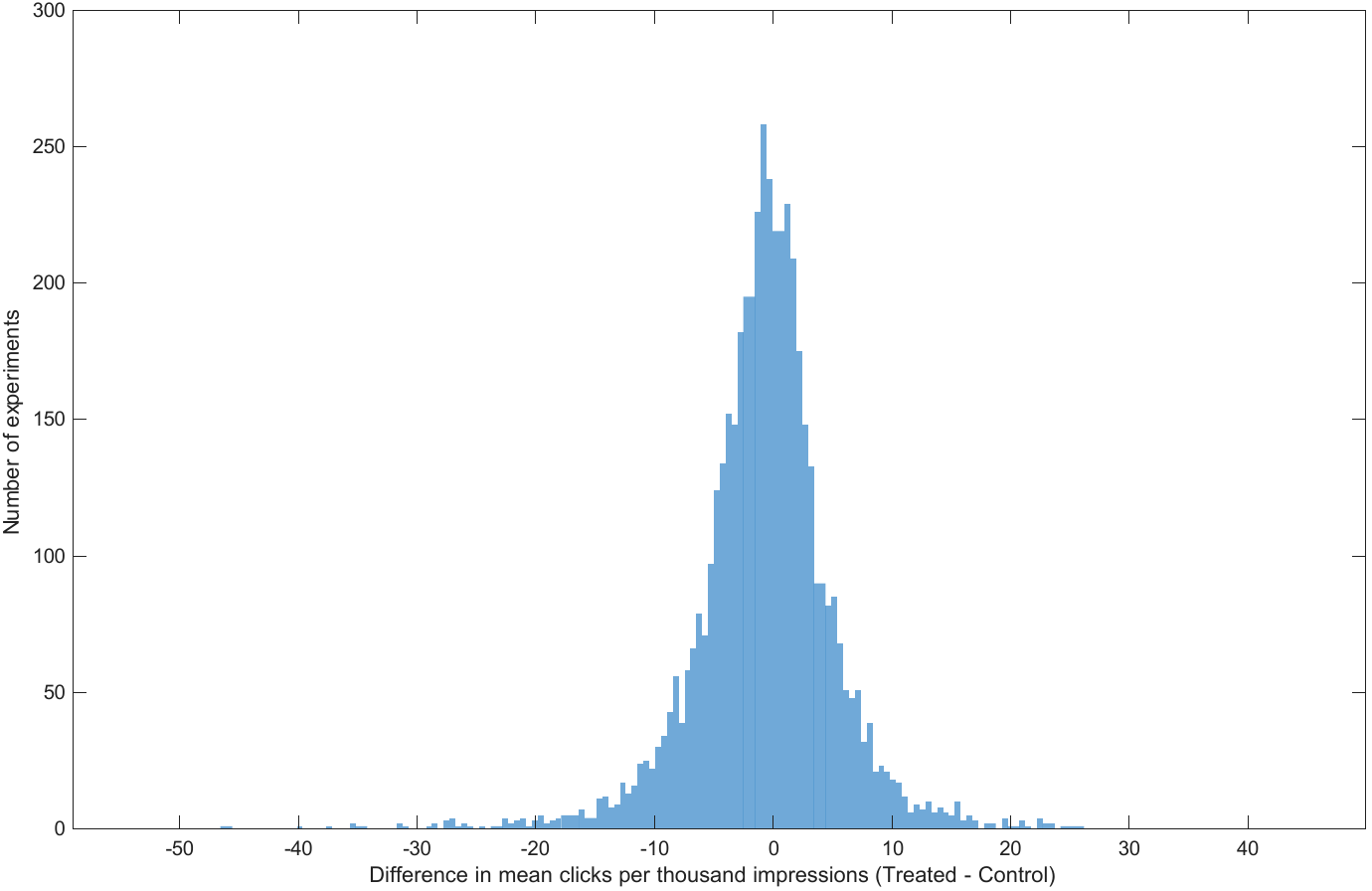}
    \label{fig:Upworthy}
\end{figure}

Figure \ref{fig:Upworthy} shows the distribution of $\widehat{\tau}$. The average value of $\widehat{\tau}$ is $-0.7621$ (clicks per thousand impressions), with range $[-54.17,,45.13]$. The standard errors have mean $2.9727$ and range $[0.2202,,7.942]$. For the remainder of this section, we use the Upworthy data to illustrate empirical Bayes estimation of the value of EBDM. Sections \ref{section:application} through \ref{section:significance} delve deeper into the EBDM value estimates for the Upworthy dataset.

\subsection{Parametric empirical Bayes}
\label{section:peB}
In this section, we adopt a Gaussian specification for the distribution of $\tau$. Because $\widehat\tau$ is unbiased, we can estimate $\mu$, the mean of the distribution of $\tau$, as the mean of $\widehat\tau$ across evaluations. To estimate $\gamma^2$, the variance of the distribution of $\tau$, we deconvolute the distribution of $\widehat\tau$ as follows. By the Total Law of Variance,
\[
\mbox{var}(\widehat\tau)=E[\mbox{var}(\widehat\tau|\tau)]+\mbox{var}(E[\widehat\tau|\tau]).
\]
Unbiasedness of $\widehat\tau$ conditional on $\tau$ implies
\begin{align*}
\gamma^2&=\mbox{var}(\tau)\\
&=\mbox{var}(E[\widehat\tau|\tau])\\
&=\mbox{var}(\widehat\tau)-E[\mbox{var}(\widehat\tau|\tau)].
\end{align*}
As a result, we define $\widehat\gamma^2$ as the difference between the variance $\widehat\tau$ across experiments in the data minus the mean of the squares of the standard errors. This estimator is not guaranteed to be non-negative \citep[see][for a discussion and alternative estimators]{morris1983parametric}. In the Upworthy dataset, $\widehat\gamma^2=36.4608-10.2437=26.2171$. 

\subsubsection{Homoskedastic case}

For the homoskedastic case, we estimate $\sigma^2$ as the average of the squares of the standard deviations of $\widehat\tau$ across studies. In the Upworthy data, this estimate is 10.4919. Plugging in this value in \eqref{equation:Vsigma} along with estimates of $\mu$ and $\gamma^2$, we obtain $V(10.2437)=1.3691$, with the value of information measured in clicks per thousand impressions.

\subsubsection{Heteroskedastic case}
\label{section:peB_het}

We now relax the assumption that $\sigma^2$ is constant. It can be shown (see appendix) that $V(\sigma^2)$ is convex. Then, by Jensen's inequality, $V(E[\sigma^2])\leq E[V(\sigma^2)]$. This result implies that the assumption of homoskedasticity may lead to underestimation of the average payoff when $\sigma^2$ is not constant. Under heteroskedasticity, we evaluate the expression in \eqref{equation:V} plugging in study-specific estimates of $\sigma^2$. Relative to the calculations in the previous section, now the value of experimentation is computed for each value of $\sigma^2$ and then integrated over the distribution of $\sigma^2$. An estimator of $V$ based on a set of policy estimates can be calculated in two steps: {\it (i)} use the square of the standard error of $\widehat\tau$ to approximate $\sigma^2$, and estimate the value of each study separately, and {\it (ii)} take the average over all studies in the sample. For the Upworthy data, this procedure yields $V=1.4057$.

\subsection{Nonparametric empirical Bayes}
We next relax the parametric restriction $\tau\sim N(\mu,\gamma^2)$ of Section \ref{section:peB} and consider a nonparametric distribution for $\tau$.  

\subsubsection{NPMLE under precision independence}
\label{section:npeB_hom}

Suppose $\tau \mid \sigma \sim G_\sigma$, where $G_\sigma$ is an unspecified distribution, and
\[
\widehat\tau \mid (\tau,\sigma) \sim N(\tau,\sigma^2).
\]
Under a precision-independence assumption (namely, that $\tau$ is independent of $\sigma$) we have
$\tau \mid \sigma \sim G_\sigma = G$, so the distribution of $\tau$ does not depend on $\sigma$.
It follows that for any $s>0$, the conditional distribution of $\tau/\sigma$ given $\sigma=s$ coincides
with the distribution of $\tau/s$.

Given $n$ studies with observed pairs
$\{(\widehat\tau_i,\sigma_i)\}_{i=1}^n$, we estimate the mixing distribution $G$ using nonparametric empirical Bayes methods. A nonparametric maximum likelihood estimator (NPMLE) of $G$ solves the problem
\begin{equation}
\max_{G\in\mathcal{G}}
\sum_{i=1}^n \log \int \phi\!\left(\frac{\widehat\tau_i-w}{\sigma_i}\right)\, dG(w),
\label{equation:npml}
\end{equation}
where $\mathcal{G}$ denotes a class of discrete distributions supported on a fixed grid
$u_1,\ldots,u_m$, with probabilities $g_1,\ldots,g_m$ (see appendix for details). Modern implementations of NPMLE (e.g.,
\citealp{koenker2017rebayes}) solve \eqref{equation:npml} efficiently and come with theoretical guarantees
\citep{jiang2020general,soloff2025multivariate}. The solution is computed over a grid $u_1, \ldots, u_m$ representing support points of $G$, with corresponding probabilities, $\widehat g_1, \ldots, \widehat g_m$.

Moreover, as shown in the appendix, the posterior mean satisfies

\begin{equation} \label{equation:posteriormean}
E[\tau|  \widehat\tau=z,\sigma=s]=\dfrac{\displaystyle \int w\phi\left(\dfrac{z-w}{s} \right)dG(w)}{\displaystyle \int \phi\left(\dfrac{z-w}{s} \right)dG(w)},    
\end{equation}

from which we derive a sample analog of $E[\tau|\widehat\tau=z,\sigma=s]$ as
\[
\frac{\displaystyle\sum_{j=1}^m u_j\phi \left(\frac{z-u_j}{s} \right)\widehat g_j}{\displaystyle\sum_{j=1}^m \phi \left(\frac{z-u_j}{s} \right)\widehat g_j}.
\]

For the Upworthy dataset, we approximate $\sigma_1,\ldots,\sigma_N$ using the reported standard errors,
estimate $G$ via the algorithm of \citet{koenker2014convex} as implemented in \citet{koenker2017rebayes},
and obtain $\widehat V=0.7621$.

\subsubsection{Relaxing the precision independence assumption}
\label{section:npeB_het}
We relax the precision-independence assumption by partitioning the range of $\widehat\sigma_1, \ldots, \widehat\sigma_n$ into five intervals and performing the NPEB calculations from the previous section within each interval. We refer to this approach as binning. Applied to the Upworthy data, binning yields $\widehat V=0.9285$.

As an alternative, we use the CLOSE-NPMLE framework of \citet{chen2024empirical}, which models the conditional distribution of the estimates given their standard errors as a flexibly parameterized location–scale family and estimates the mixing distribution nonparametrically via NPMLE. For the Upworthy data, CLOSE-NPMLE yields $\widehat V=0.9560$.

\section{Simulations}
\label{section:Simulations}

We consider two data generating processes (DGP). In DGP1, the parameters $\tau$ have a standard Gaussian distribution $\tau\sim N(0,1)$, so the parametric empirical Bayes model of Section \ref{section:gaussian_prior} applies. In DGP2, the parameters $\tau$ follow the mixture distribution as in Section \ref{section:mixture_prior}. In particular, in DGP2, 
\[
\tau \sim \left\{\begin{array}{ll}N(-5,1/2)&\mbox{with prob. } 0.01,\\
N(0,1/2)&\mbox{with prob. } 0.98,\\
N(5,1/2)&\mbox{with prob. } 0.01.\\\end{array}\right.
\]
DGP1 and DGP2 both produce a distribution of $\tau$ with mean zero and variance one. We generate $\widehat\tau$ as $\widehat\tau=\tau + \sigma u$, where $u$ is independent standard Gaussian and $\sigma=0.1$. To calculate the expected payoff of EBDM, we consider the case of $c_L=0$. 

We run 1000 simulations for DGP1 and DGP2 with $n=500$. Equation \eqref{equation:Vsigma} with $\mu=0$, $\gamma^2=1$, $\sigma=0.1$, and $c_L=0$ gives the true expected payoff of EBDM under DGP1. To calculate the true expected payoff of EBDM under DGP2, 
we first use equation \eqref{equation:mixture_posterior_mean} to compute $E[\tau|\widehat\tau]$ over the $n\times 1000=500{,}000$ realizations of $\widehat\tau$ in the simulations, and report the average of $\max\{E[\tau|\widehat\tau],0\}$. In each of the simulations, we calculate parametric and nonparametric empirical Bayes estimates of the average payoff of EBDM. The parametric empirical Bayes estimator is the sample analog of equation \eqref{equation:V}. This estimator is valid under the assumption that the true distribution of $\tau$ is Gaussian. The nonparametric empirical Bayes estimator is as in Section \ref{section:npeB_hom}. For the simulations in this section, we treat $\sigma^2$ as known.

\begin{table}[!tp]
\caption{Simulation results}
\centering
\begin{tabular}{lccccc}\\[-1ex]\hline\hline\\[-2ex]
                    &&&&\multicolumn{2}{c}{empirical Bayes}\\
                    \hspace*{-0.2cm}\raisebox{-1ex}{distribution of $\tau$:}&&\hspace*{2cm}true value          &&parametric &nonparametric\\ \cline{5-6}\\[-2ex]
%\hspace*{.15cm}
\multirow{2}{*}{\begin{tabular}{l}\hspace*{1.28cm}{Gaussian}\ $\begin{flcases}\mbox{expected payoff}\\ \mbox{95\% interval}\end{flcases}$\end{tabular}}    &&\hspace*{2cm}0.3970&     &0.3962        &0.3963\\
&&&&$[0.3450,0.4475]$&$[0.3443,0.4483]$\\[2ex]
\multirow{2}{*}{\begin{tabular}{c}\hspace*{1.5cm}{mixture}\ $\begin{flcases}\mbox{expected payoff}\\ $\mbox{95\% interval}$\end{flcases}$\end{tabular}}    &&\hspace*{2cm}0.3230     &&0.3939       &0.3232\\
&&&&$[0.3121,0.4757]$&$[0.2645,0.3819]$\\[1ex] \hline
\end{tabular}
\label{table:simulations}
\end{table}

Table \ref{table:simulations} reports the true values of the expected payoff of EBDM along with means and 95 percent intervals for the distribution of the estimates across simulations. When $\tau$ is Gaussian, the distributions of the parametric and nonparametric empirical Bayes estimates across simulations are both centered near the true value of the expected EBDM payoff. Moreover, there is no evidence of substantial gains from knowledge of the parametric form of the distribution of $\tau$. The 95 percent interval for the nonparametric estimator is only 1.5 percent wider than the interval for the parametric estimator.   

For the case when the distribution of $\tau$ is a mixture, the results for the parametric estimator reveal a clear bias, while the distribution of the nonparametric estimator remains centered at the true value of the expected payoff. Moreover, the 95 percent interval for the nonparametric estimator is 28.3 percent narrower than the interval for the parametric estimator.   

\section{Application to the Upworthy dataset}
\label{section:application}

Table \ref{table:upworthy} reports parametric and nonparametric empirical Bayes estimates of the value of EBDM in the Upworthy data. In the parametric case, the table reports estimates computed under heteroskedasticity (Section \ref{section:peB_het}). In the nonparametric case, it reports three estimates: the precision-independence NPMLE (Section \ref{section:npeB_hom}), and binning and CLOSE-NPMLE estimates (Section \ref{section:npeB_het}) that relax the assumption of precision independence. Below each of the estimates of the value of EBDM, Table \ref{table:upworthy} reports 95 percent intervals computed over $1{,}000$ bootstrap draws from the distribution of $(\widehat\tau,\widehat\sigma^2)$ in the data. In our calculations, we impose $c_L=c_F=c(\sigma^2)=0$.  

Because $\widehat\tau$ has a negative mean and large dispersion relative to its mean, this is a setting where we expect to have substantial gains from EBDM. Indeed, the parametric model suggests a $\text{VoE}$ of about $1.8$ times the magnitude of $\widehat\mu$ (but with a positive sign), while nonparametric empirical Bayes under the most restrictive specification yields a value only slightly larger than $\widehat{\mu}$ in magnitude. The most flexible specifications (binning and CLOSE) deliver similar results, implying a $\text{VoE}$ that is 25.4 percent larger than $\widehat{\mu}$ in magnitude and with a positive sign and a $\text{VoID}$ that is 125.4 percent larger than $\widehat{\mu}$ in magnitude, again with a positive sign.

\begin{table}[!tp]
\caption{Value of EBDM for the Upworthy data}
\vspace{0.1cm}   % <- add this
\label{table:upworthy}
\centering
\begin{threeparttable}
\begin{tabular}{lcccccc}\\[-1ex]\hline\hline\\[-2ex]
{\it Distribution of $\widehat\tau$:}\\
\multicolumn{7}{c}{$\widehat\mu = -0.7622$}\\
\multicolumn{7}{c}{(standard deviation of $\widehat\tau$ is 6.03)}\\[1ex]
{\it Value of EBDM:}\\
                    &&parametric &&\multicolumn{3}{c}{nonparametric}\\
                    \cline{5-7}\\[-2ex]
                    &&    &&precision  &binning  & CLOSE-NPMLE\\
                    &&     && independence &  & \\
                     && (1) && (2)  & (3) & (4) \\                   
                    \cline{5-7}\\[-2ex]
\multirow{2}{*}{\begin{tabular}{l}$\text{VoE}$\hspace*{.3cm}$\begin{flcases}\mbox{estimate}\\ \mbox{95\% interval}\end{flcases}$\end{tabular}}    &&  1.4057    && 0.7621 & 0.9285  & 0.9560   \\
&&$[1.2792,1.5321]$&&$[0.6743,0.8499]$ & $[0.8290,1.0280]$ &$[0.8519,1.0601]$\\[1ex]
\multirow{2}{*}{\begin{tabular}{c}$\text{VoID}$\ $\begin{flcases}\mbox{estimate}\\ \mbox{95\% interval}\end{flcases}$\end{tabular}}    && 2.1678       &&   1.5242 & 1.6907  & 1.7181 \\
&&$[1.9906,2.3451]$&&$[1.3832,1.6653]$ & $[1.5495,1.8319]$ & $[1.5794,1.8568]$\\[1ex] \hline
\end{tabular}
\begin{tablenotes}
\small\smallskip
\hspace*{-0.6cm}\begin{minipage}{1.05\textwidth}
\item \textit{Note:}~The table reports estimates of the $\text{VoE}$ and $\text{VoID}$ for the Upworthy data $(n=4{,}857)$. For each method, the table presents the point estimate of the payoff and a dispersion interval obtained from $1{,}000$ bootstrap samples. The parametric model is computed under a heteroskedasticity assumption.
\end{minipage}
\end{tablenotes}
\end{threeparttable}
\end{table}

\section{Estimation of counterfactual EBDM values}
\label{section:counterfactual}

This section provides estimates of the value of EBDM under alternative levels of statistical precision and under alternative levels of dispersion in the distribution of the effects of the policies.

First, we estimate how the value of EBDM would change as a result of a change in $\sigma^2$. In the resulting counterfactuals, the variance of the estimators is equal to the variance of $\widehat\tau_1\,|\,\tau_1, \ldots, \widehat\tau_n\,|\,\tau_n$ in the original sample multiplied by $\lambda$. That is, $\lambda=0.5$ represents a counterfactual scenario where the variances of the estimators are 50 percent smaller than the variance estimates in the original sample, while for $\lambda=1.5$ the variances of the estimators are 50 percent larger than in the original sample. 

For simplicity, we consider only counterfactual scenarios such that $\tau$ is independent of estimation variance, $\sigma^2$, and estimate the value of EBDM using the parametric empirical Bayes estimator of Section \ref{section:peB}. It is conceptually straightforward to extend this procedure to more general settings (e.g., by modeling the dependence between $\tau$ and $\sigma^2$ and/or using nonparametric empirical Bayes estimators). 

For each estimate $i=1, \ldots, n$ in our sample, we draw a value from the empirical Bayes estimate of the distribution of $\tau$. Let $\tau_1^*, \ldots , \tau_n^*$ be the resulting values for the draws. Next, for $i=1, \ldots, n$, we obtain $\widehat\tau_i^*=\tau_i^*+\sigma_i^* U_i$, where $U_1, \ldots, U_n$ are independent draws from the standard Gaussian distribution, and $\sigma_i^*=\sqrt{\lambda}\widehat\sigma_i$. We use the new sample $(\widehat\tau_1^*,\widehat\sigma_1^*), \ldots, (\widehat\tau_n^*,\widehat\sigma_n^*)$ to compute an estimate of the value of EBDM. We repeat this procedure multiple times to obtain the distribution of EBDM-value estimates for a particular value of $\lambda$. The average of this distribution is our estimate of the value of EBDM under variance modification factor $\lambda$. For the parametric empirical Bayes case, this average can also be computed directly using an empirical counterpart of equation \eqref{equation:V} that applies the variance modification factor $\lambda$ to $\sigma^2$.

\begin{figure}[!tp]
    \centering
    \caption{Counterfactual values of EBDM}
    \includegraphics[width=\linewidth]{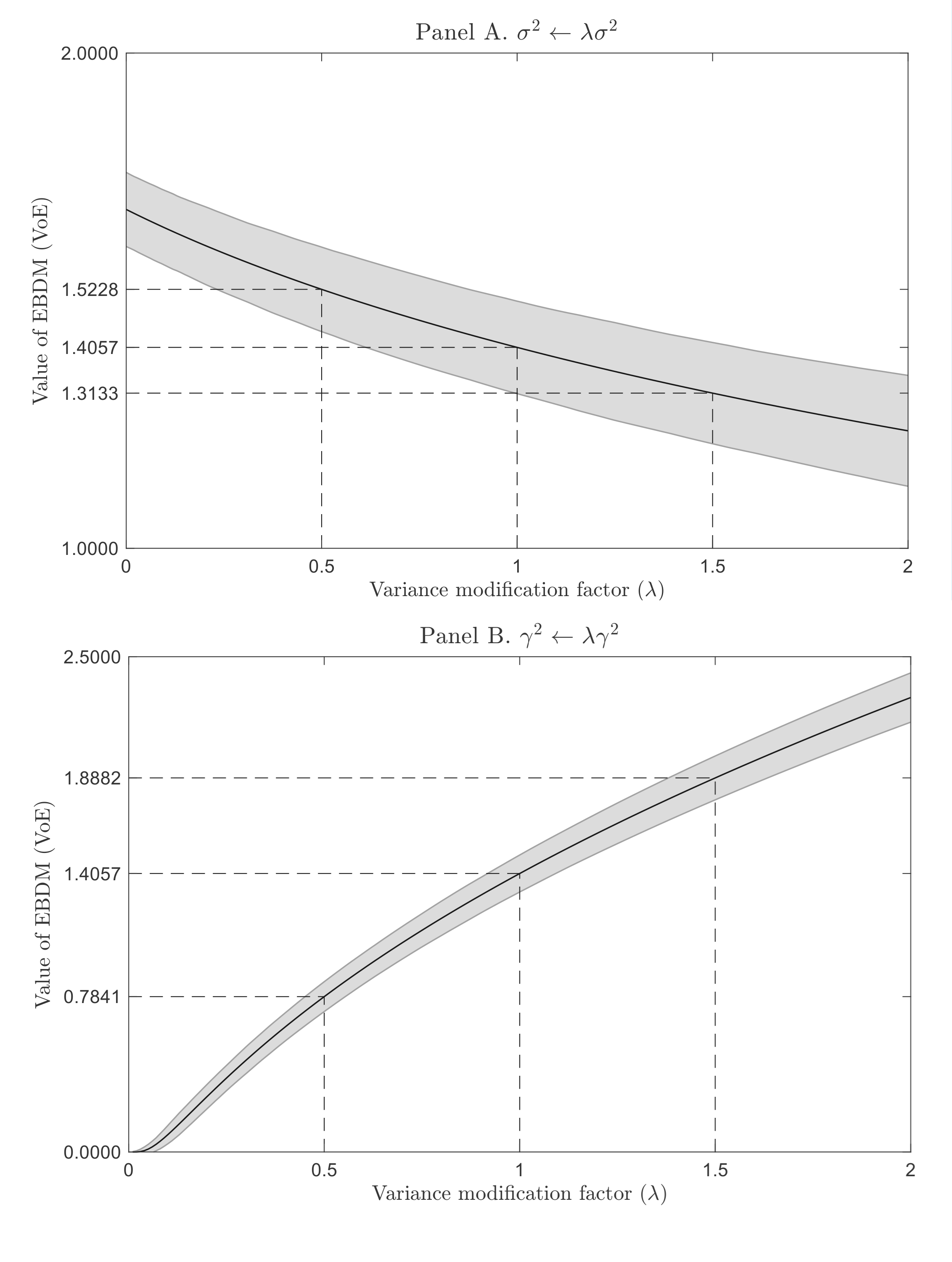}
    \label{fig:counterfactual_upworthy}
\end{figure}

Panel A of Figure \ref{fig:counterfactual_upworthy} reports the results obtained from applying the procedure described above to the Upworthy data. The solid line represents the value of EBDM as a function of the variance modification factor, $\lambda$. The shaded area represents 95 percent intervals from the distribution of EBDM estimates. An investment that reduces estimation variance by half (about a 29.3 percent decrease in standard errors) leads to an increase in the value of EBDM by $8.33$ percent, from $1.4057$ to $1.5228$. Conversely, an increase in estimation variance by half (about a 22.5 percent increase in standard errors) decreases the value of EBDM by 6.56 percent, from $1.4057$ to $1.3133$.

We next compute counterfactual EBDM values for different levels of $\gamma^2$, the variance of the distribution of $\tau$. For each observation $i = 1, \dots, N$ in our sample, we draw a value from a distribution with the same mean as the empirical Bayes estimate of the distribution of $\tau$ but with variance adjusted by a factor $\lambda$. Let $\tau_1^*, \dots, \tau_n^*$ denote these draws. 
Next, for each $i$, we compute $\widehat\tau_i^* = \tau_i^* + \widehat\sigma_i U_i$, where $U_1, \ldots, U_n$ are independent standard Gaussian draws. 
Using the sample $(\widehat\tau_1^*, \widehat\sigma_1), \dots, (\widehat\tau_n^*, \widehat\sigma_n)$, we estimate the value of EBDM. Repeating this procedure multiple times yields a distribution of EBDM estimates for a given $\lambda$. The mean of this distribution is our estimate of EBDM under variance modification factor $\lambda$ for $\gamma^2$.

Panel B of Figure \ref{fig:counterfactual_upworthy} shows how changes in the heterogeneity of true policy effects influence the value of EBDM. 
When the variance modification factor $\lambda$ is greater than one, meaning that the variance of the true effects increases, the value of EBDM rises. For instance, when $\gamma^2$ is increased by 50 percent ($\lambda = 1.5$), the estimated value of EBDM increases from $1.4057$ to $1.8882$, reflecting a 34.32 percent gain. Conversely, when $\gamma^2$ is reduced by half ($\lambda = 0.5$), the value of EBDM declines to $0.7841$, representing a 44.22 percent decrease from the baseline.

\section{Significance testing reduces the value of information}\label{section:significance}

Organizations commonly implement interventions only when they {\em (i)} deliver positive estimated treatment effects and {\em (ii)} attain statistical significance at a prespecified level. This section shows that significance-based decision rules fail to exploit the full informational content of the data and are therefore suboptimal from an EBDM perspective.

Intuitively, statistical significance decision rules prioritize Type I error control rather than maximizing expected payoff. Therefore, they may discard interventions with large but imprecisely estimated effects that could generate substantial rewards, or systematically favor interventions with small but precisely estimated effects, thereby biasing decisions toward low-variance interventions rather than those with high expected payoffs. Moreover, conditioning implementation decisions on statistical significance induces a winner's curse effect: selected interventions are disproportionately likely to be those whose effects were overestimated in the sample due to sampling variation \citep[see][]{andrews2023inference}. Empirical Bayes methods address these challenges by shrinking extreme estimates toward the prior distribution of treatment effects, thereby removing substantial noise from signals. This enables decision-makers to better distinguish signal from noise and rank interventions according to their posterior expected value, thus favoring interventions with genuinely positive expected payoffs.

Define the value of EBDM under a statistical significance rule with one-sided significance level $\alpha$ as
\begin{align*}V_{\text{sig}}&=E[(\tau  - c_L)  I_{(z_{1-\alpha},\infty)}((\widehat\tau-c_L)/\sigma)]\\
&=E[E[(\tau  - c_L) \,|\, \widehat\tau, \sigma] I_{(z_{1-\alpha},\infty)}((\widehat\tau-c_L)/\sigma)]
\end{align*}
where $z_{1-\alpha}$ is the $(1-\alpha)$-th quantile of a standard normal distribution.

Consider first the case of a fixed value of $\sigma$. For the parametric model $\tau\sim N(\mu,\gamma^2)$, the marginal distribution of $\widehat\tau$ is $\widehat\tau \,|\, \sigma \sim N(\mu, \gamma^2 + \sigma^2)$. As a result, conditional on $\sigma^2$, the value of EBDM under a statistical significance rule at significance level $\alpha$ is
\begin{align*}
V_{\text{sig}}(\sigma^2) & = 
E[(\tau  - c_L)  I_{(z_{1-\alpha},\infty)}((\widehat\tau-c_L)/\sigma)|\sigma]\\
&=E \left[ \left(\frac{\mu / \gamma^2 + \widehat\tau / \sigma^2}{1 / \gamma^2 + 1 /  \sigma^2}-c_L\right)I_{(z_{1-\alpha},\infty)}((\widehat\tau-c_L)/\sigma) \; \Big | \; \sigma\right].
\end{align*}
Calculations in the appendix show

\begin{equation}\label{equation:closedsignificance}
 V_{\text{sig}}(\sigma^2) = 
(\mu - c_L) \Phi\left(\frac{(\mu-c_L)-z_{1-\alpha}\sigma}{\sqrt{\gamma^2+\sigma^2}}\right)
+ \frac{\gamma^2}{\sqrt{\gamma^2+\sigma^2}}\phi\left(\frac{(\mu-c_L)-z_{1-\alpha}\sigma}{\sqrt{\gamma^2+\sigma^2}}\right).   
\end{equation}

Compare this expression to $V(\sigma^2)$ in \eqref{equation:Vsigma}. It holds that 
\begin{equation}
\label{equation:optsig}
V_{\text{sig}}(\sigma^2)\leq V(\sigma^2),
\end{equation}
provided $\gamma^2>0$, with strict inequality except for the case in which the arguments of the functions $\Phi(\cdot)$ and $\phi(\cdot)$ coincide in the expression of $V(\sigma)$ and $V_{\text{sig}}(\sigma)$. The appendix contains a detailed comparison of the payoffs $V(\sigma)$ and $V_{\text{sig}}(\sigma)$.

A plug-in procedure yields the estimator of $V_{\text{sig}}$
\[\dfrac{1}{n} \sum \limits_{i=1}^n\left[(\widehat{\mu} - c_L) \Phi\left(\frac{\widehat{\mu} - c_L-z_{1-\alpha}\sigma_i}{\sqrt{\widehat{\gamma}^2+\sigma_i^2}} \right) + \frac{\widehat{\gamma}^2}{\sqrt{\widehat{\gamma}^2+\sigma_i^2}}\phi\left(\frac{\widehat{\mu} - c_L-z_{1-\alpha}\sigma_i}{\sqrt{\widehat{\gamma}^2+\sigma_i^2}}\right)\right].\]

As an alternative, for any estimator of the posterior mean $\widehat{E}[\tau \,|\, \widehat{\tau}_i,\sigma_i]$ (parametric or nonparametric)
an estimator of the value of EBDM under a statistical significance decision rule is given by
\begin{align*}
\dfrac{1}{n} \sum \limits_{i=1}^n \left(\widehat{E}[\tau \mid \widehat{\tau}_i,\sigma_i]-c_L \right) I_{(z_{1-\alpha},\infty)}((\widehat\tau_i-c_L)/\sigma_i). 
\end{align*}

\begin{table}[!tp] 
\caption{The value of optimal vs.~significance decision rules in the Upworthy data}
\vspace{0.1cm} % <- add this 
\label{table:upworthy_results_sig} 
\centering \begin{threeparttable} 
\begin{tabular}{lcccccc}\\[-1ex]
\hline\hline\\[-2ex] 
                    &&parametric &&\multicolumn{3}{c}{nonparametric}\\
                    \cline{5-7}\\[-2ex]
                    &&    &&precision  &binning  & CLOSE-NPMLE\\
                    &&     && independence &  & \\
                     && (1) && (2)  & (3) & (4) \\                   
                    \cline{5-7}\\[-2ex]
%\hspace*{.15cm} 
$\text{VoE}$ &&  1.4057    && 0.7621 &0.9285  &0.9560   \\
Significance rule ($\alpha=0.05$) && 1.0287 && 0.5910 & 0.6510 & 0.6609 \\
\hline
\end{tabular} 
\begin{tablenotes}
\small\smallskip
\hspace*{-0.6cm}\begin{minipage}{1.0\textwidth}
\item \textit{Note:}~The table reports estimates of the $\text{VoE}$ and the value under a 5 percent significance decision rule for the Upworthy data $(n=4{,}857)$. The parametric estimates are for the heteroskedastic case (Section \ref{section:peB_het}).
\end{minipage}
\end{tablenotes}
\end{threeparttable}
\end{table}

Table \ref{table:upworthy_results_sig} compares the value delivered by significance-based EBDM to the $\text{VoE}$ for the Upworthy data. Across specifications, the empirical Bayes procedures in Section \ref{section:empirical_Bayes} deliver substantially higher value than significance-based decision making. Under the parametric heteroskedastic model, EBDM yields $1.4057$, compared with $1.0287$ under the 5 percent significance rule—a reduction of about 27 percent. The gap is larger under nonparametric methods: binning and CLOSE yield values between $0.93$ and $0.95$, whereas the corresponding significance-rule values cluster around $0.66$–$0.67$, implying that significance screening discards roughly 30 percent of attainable value. Overall, these results show that significance-oriented decision rules systematically underperform value-based policies, especially in settings with substantial heterogeneity and estimation noise, where empirical Bayes methods can extract value from interventions that significance tests would discard.

\section{Conclusions}

This article develops an empirical framework to quantify the value of evidence-based decision making and to examine how statistical precision and heterogeneity in policy effects moderate that value. Using both parametric and nonparametric empirical Bayes methods, we estimate the benefit of incorporating data-driven evidence into decision-making processes by balancing the trade-off between the costs of acquiring information and the expected improvements in outcomes.
Higher statistical precision (i.e., lower $\sigma^2$) and greater heterogeneity in policy effects (i.e., higher $\gamma^2$) increase the value of evidence-based decision making.

Our framework provides a principled approach for organizations to evaluate whether investing in additional data collection or policy exploration is worthwhile given the expected gains in decision quality. The proposed methods are particularly relevant for organizations that frequently conduct experiments to optimize policies and business strategies. The availability of many experimental evaluations---that is, the availability of many instances of $(\widehat\tau, \widehat\sigma^2)$---makes it possible to estimate the distribution of policy effects. 

Future research could extend our framework by incorporating more flexible empirical Bayes estimators and exploring settings where estimation variance and treatment effect heterogeneity are determined endogenously by prior information. Additionally, applying our approach to firm-level, governmental, and healthcare decision-making contexts could further validate its usefulness in diverse policy environments.

Currently, many organizations rely on power calculations to guide their study designs, without taking into account the cost-benefit trade-offs associated with evidence-based decision making. As organizations continue to expand their reliance on data for policy decisions, our proposed methods offer a practical procedure for optimizing information acquisition strategies. 

\bigskip

\centerline{\bf Appendix}
\appendix

\underline{\textit{Convexity of $V(\sigma)$:}} The second derivative of $V(\sigma^2)$ with respect to $\sigma^2$ is
\begin{equation}
    \frac{\partial^2 V(\sigma^2)}{\partial \sigma^2\partial\sigma^2} =
    \frac{3\gamma^4 + (\mu-c_L)^2(\gamma^2+\sigma^2)}{4\gamma^2(\gamma^2+\sigma^2)^{5/2}}\phi\left(\frac{\mu-c_L}{\gamma^2/\sqrt{\gamma^2+\sigma^2}}\right)\geq 0. 
\end{equation}
Now, Jensen's inequality implies $V(E[\sigma^2])\leq E[V(\sigma^2)]$.
\bigskip

\underline{\textit{Derivation of \eqref{equation:npml}:}} 
For any integrable function $h$ and any $s>0$,
\[
\int h(t)\, dF_{\tau/\sigma \mid \sigma}(t \mid s)
=E[h(\tau/\sigma)\mid \sigma=s]
=E[h(\tau/s)]
=\int h(w/s)\, dG(w).
\] 
By the law of total probability,
\begin{align*}
f_{\widehat{\tau}/\sigma \mid \sigma}(r \mid s)
&=\int f_{\widehat{\tau}/\sigma \mid \tau/\sigma,\sigma}(r \mid t,s)\,
dF_{\tau/\sigma \mid \sigma}(t \mid s)\\
&=\int \phi(r-t)\, dF_{\tau/\sigma \mid \sigma}(t \mid s)\\
&=\int \phi\!\left(r-\frac{w}{s}\right)\, dG(w).
\end{align*}

Now consider $n$ studies indexed by $i=1,\ldots,n$, with observed pairs
$\{(\widehat\tau_i,\sigma_i)\}_{i=1}^n$. Then, the likelihood of the data $\{(\widehat{\tau}_i,\sigma_i)\}_{i=1}^n$ is
\[\prod_{i=1}^n f_{\widehat{\tau}/\sigma \mid \sigma} \left( \frac{\widehat{\tau}_i}{\sigma_i} \;\Big \vert \; \sigma_i \right)= \prod_{i=1}^n \int_{\mathbb
R} \phi\left(\frac{\widehat{\tau}_i-w}{\sigma_i}\right)dG(w).\]

justifying the nonparametric Maximum Likelihood procedure.

\bigskip

\underline{\textit{Proof of \eqref{equation:posteriormean}:}} Since $\widehat{\tau} \mid (\tau=w, \sigma=s) \sim N(w,s^2)$, it follows that $f_{\widehat{\tau} \mid \tau, \sigma}(z \mid w, s)=\phi((z-w)/s)/s$. Now, the law of total probability implies
\[f_{\widehat{\tau} \mid \sigma}(z \mid s)=\int f_{\widehat{\tau} \mid \tau,\sigma}(z \mid w,s)dG(w)= \dfrac{1}{s}\int \phi\left(\dfrac{z-w}{s} \right)dG(w). \]

Moreover, it follows from Bayes' rule that
\begin{align*}dF_{\tau \mid \widehat{\tau},\sigma}(w \mid z,s)&=\dfrac{f_{\widehat{\tau} \mid \tau,\sigma}(z \mid w,s) dG(w)}{f_{\widehat{\tau} \mid \sigma}(z \mid s)}\\
&=\dfrac{\phi\left(\displaystyle\frac{z-w}{s}\right)dG(w)}{\displaystyle\int\phi\left(\frac{z-u}{s}\right)dG(u)}
\end{align*}
whereby,
\[E[\tau|  \widehat\tau=z,\sigma=s]=\int wdF_{\tau \mid \widehat{\tau},\sigma}(w \mid z,s)=\dfrac{\displaystyle \int w\phi\left(\dfrac{z-w}{s} \right)dG(w)}{\displaystyle \int \phi\left(\dfrac{z-w}{s} \right)dG(w)}.\]

\bigskip

\underline{\textit{Proof of \eqref{equation:closedsignificance}:}}  For the parametric model $\tau\sim N(\mu,\gamma^2)$, the marginal distribution of $\widehat\tau$ is $\widehat\tau \,|\, \sigma \sim N(\mu, \gamma^2 + \sigma^2)$. As a result, the value of EBDM under a statistical significance rule at significance level $\alpha$ is
\begin{align*}
V_{\text{sig}}(\sigma^2)=E \left[ \left(\frac{\mu / \gamma^2 + \widehat\tau / \sigma^2}{1 / \gamma^2 + 1 /  \sigma^2}-c_L\right)I_{(z_{1-\alpha},\infty)}((\widehat\tau-c_L)/\sigma) \; \Big | \; \sigma\right].
\end{align*}

Let $\widehat t = (\widehat\tau - c_L)/ \sigma$, so $\widehat t \mid \sigma \sim N((\mu-c_L)/\sigma,(\gamma^2+\sigma^2)/\sigma^2)$. Then,
\[ V_{\text{sig}}(\sigma) =(\mu-c_L) \Pr(\widehat t>z_{1-\alpha} \mid \sigma)+\left( \dfrac{\gamma^2}{\gamma^2+\sigma^2} \right)E[ (\widehat{\tau}-\mu)I_{(z_{1-\alpha}, \infty)}(\,\widehat t\,) \mid \sigma]. \]

Note that
\begin{align*}
\Pr(\widehat t>z_{1-\alpha}\mid \sigma)&=\Pr\left(\frac{\widehat t - (\mu-c_L)/\sigma}{\sqrt{(\gamma^2+\sigma^2)/\sigma^2}}> \frac{z_{1-\alpha} - (\mu-c_L)/\sigma}{\sqrt{(\gamma^2+\sigma^2)/\sigma^2}} \; \Bigg \vert \; \sigma\right)\\
& = 1- \Phi\left(\frac{z_{1-\alpha} - (\mu-c_L)/\sigma}{\sqrt{(\gamma^2+\sigma^2)/\sigma^2}}\right).
\end{align*}

Moreover, it follows from integration by parts that
\begin{align*}
E[ (\widehat{\tau}-\mu)I_{(z_{1-\alpha},\infty)}(\,\widehat t\,) \mid \sigma]&= \int_{c_L+z_{1-\alpha}\sigma}^\infty (t-\mu)\dfrac{1}{\sqrt{\gamma^2+\sigma^2}} \phi\left(  \dfrac{t-\mu}{\sqrt{\gamma^2+\sigma^2}}\right)dt  \\
& = \sqrt{\gamma^2+\sigma^2} \phi \left( \dfrac{c_L+z_{1-\alpha} \sigma-\mu}{\sqrt{\gamma^2+\sigma^2}} \right).    
\end{align*}

Therefore,
\begin{align*}
V_{\text{sig}}(\sigma) &= (\mu - c_L)\left(1- \Phi\left(\frac{z_{1-\alpha} - (\mu-c_L)/\sigma}{\sqrt{(\gamma^2+\sigma^2)/\sigma^2}}\right)\right)
+ \frac{\gamma^2}{\sqrt{\gamma^2+\sigma^2}}\phi\left(\displaystyle\frac{z_{1-\alpha} - (\mu-c_L)/\sigma}{\sqrt{(\gamma^2+\sigma^2)/\sigma^2}}\right)\\
&= (\mu - c_L) \Phi\left(\frac{(\mu-c_L)-z_{1-\alpha}\sigma}{\sqrt{\gamma^2+\sigma^2}}\right)
+ \frac{\gamma^2}{\sqrt{\gamma^2+\sigma^2}}\phi\left(\frac{(\mu-c_L)-z_{1-\alpha}\sigma}{\sqrt{\gamma^2+\sigma^2}}\right).
\end{align*}

\bigskip

\underline{\textit{Proof of \eqref{equation:optsig}:}} Consider the function
$f(x) = a \Phi(x) + b \phi(x)$,
with $b>0$. Using the fact that $\partial \phi(x)/\partial x = -x\phi(x)$, it follows that
\[
\frac{\partial f(x)}{\partial x} = (a   - b x) \phi(x).
\]
That is, $\partial f(x)/\partial x$ is positive for $x<a/b$, equal to zero for $x=a/b$, and negative for $x>a/b$. As a result, $x^* = a/b$ gives the unique global maximum. Making $a=\mu-c_L$ and $b = \gamma^2/\sqrt{\gamma^2+\sigma^2}$ proves \eqref{equation:optsig}. 

\bigskip

\underline{\textit{Comparison of  $V(\sigma)$ and $V_{\text{sig}}(\sigma)$}}

First, notice that $V_{\text{sig}}(\sigma^2)$ and $V(\sigma^2)$ are continuous in $\sigma^2$, and $V_{\text{sig}}(0)=V(0)$. That is, the two rules extract similar values in low-noise environments. 

As the scale of the noise, $\sigma^2$, increases, both $V_{\text{sig}}(\sigma^2)$ and $V(\sigma^2)$ decrease. Indeed, some algebra shows, 
\[
\frac{\partial V_{\text{sig}}(\sigma^2)}{\partial \sigma^2} = 
-\left(\frac{\gamma^2(\gamma^2+\sigma^2)+((\mu-c_L)\sigma+z_{1-\alpha}\gamma^2)^2}{2(\gamma^2+\sigma^2)^{5/2}}\right)
\phi\left(\frac{(\mu-c_L)-z_{1-\alpha}\sigma}{\sqrt{\gamma^2+\sigma^2}}\right)<0.
\]
Moreover
\[
\lim_{\sigma^2\rightarrow \infty}
V_{\text{sig}}(\sigma^2) = (\mu-c_L)\,\alpha,
\] and
\[
\lim_{\sigma^2\rightarrow \infty} V(\sigma^2) = \max\{\mu-c_L,0\}.
\]
That is, as $\sigma^2$ increases, significance testing loses power and rejects with probability $\alpha$. At the same time, as  $\sigma^2$ increases, $\widehat\tau$ loses informativeness and $E[\tau\,|\,\widehat\tau, \sigma^2]$ converges to $\mu$. This explains the limit of $V_{\text{sig}}(\sigma^2)$. For sufficiently large $\sigma^2$, the value of EBDM under a significance decision rule is negative if $\mu-c_L<0$, and recovers only a small fraction $\alpha$ of $\mu-c_L$ if $\mu-c_L>0$. In contrast, $V(\sigma^2)$ is always non-negative and always recovers at least the full value of $\mu-c_L$ when $\mu-c_L>0$.  

\newpage

\bibliography{bib}

\end{document}